\documentclass[preprint,aps,floats,showpacs]{revtex4}
 
\usepackage{epsf}

\begin{document}

\title{Constraints on proton-proton fusion from helioseismology}
\author{K.I.T. Brown, M.N. Butler, and D.B. Guenther}
\affiliation{Department of Astronomy and Physics, Saint Mary's University\\
Halifax, NS B3H 3C3 Canada }
 
\begin{abstract}
 
 The proton-proton ($pp$) fusion cross-section found at the heart of solar models
 is unconstrained experimentally and relies solely on theoretical calculations.
 Effective field theory provides an opportunity to constrain the $pp$ cross-section
 experimentally, however, this method is complicated by the appearance of two-nucleon effects
 in the form of an unknown parameter $L_{1,A}$.  We present a method to constrain $L_{1,A}$
 using the Standard Solar Model and helioseismology.  Using this method, we determine
 a value of $L_{1,A}$ = 7.0 fm$^{3}$ with a range of 1.6 to 12.4 fm$^{3}$.  These results
 are consistent with theoretical estimates of $L_{1,A} \approx$ 6 fm$^{3}$. 
 
\end{abstract}

\maketitle

\vfill\eject

\section{Introduction}

The Standard Solar Model (SSM) is a model of the sun constructed by numerically solving the
equations of stellar structure and evolution constrained by the observed composition, luminosity,
and radius of the sun at the current age. This model yields acoustic oscillation frequencies ($p$-modes)
that match the observed $p$-mode oscillation spectrum of the sun to better than 0.3\%~\cite{Gue97}. The uncertainties
associated with the observed oscillation frequencies are an order of magnitude smaller.  The excellent 
agreement between the model and observed frequency spectra implies that the run of sound speed predicted
by the model is close to that of the sun which, in turn, suggests that the physics used to construct
the model of the sun is accurate.

The individual $p$-mode frequencies extend to distinct depths in the sun, and hence can be used to study
the physics of specific regions.  Of interest to the work described here are the low-$l$ (where
$l$ corresponds, in spherical harmonic nomenclature, to the number of nodes in azimuth) $p$-mode
frequencies that penetrate deep into the core of the sun.  These modes, grouped in combinations that
cancel out surface dependencies, can be used to test the important physics of the core.  One of the greatest
sources of uncertainty in the physics of the solar model core is the $pp$-fusion cross-section.  The reaction
is unconstrained by laboratory-based experiments, relying completely on precise calculations from theoretical
nuclear physics.  Owing to the precise agreement between the observed and predicted oscillation frequencies
that already exists, it is possible to use the oscillation frequencies to constrain the $pp$ cross-section, that
is, to determine the range of cross-sections that yield oscillation frequencies within the known uncertainties
of the solar model physics and constraints.  To date, this sort of constraint has
been imposed only through simplifications of the standard solar 
model~\cite{antia}, and here we use a fully-developed stellar evolution code, to
be discussed later.

\begin{figure}[t]
\centerline{{\epsfxsize=5.0in \epsfbox{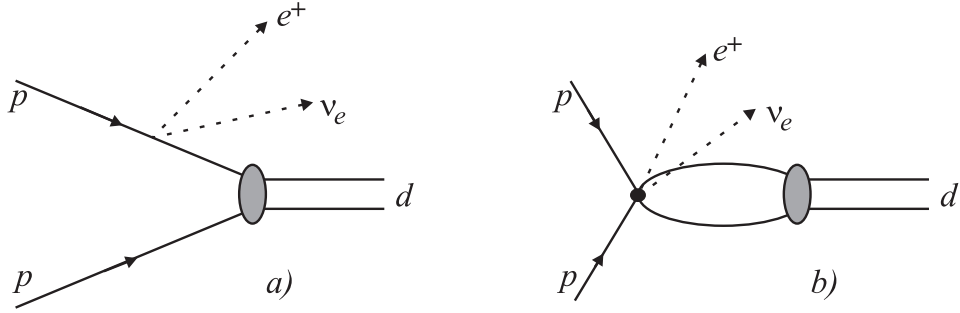}}}
\noindent
\caption{{\it The process $pp\rightarrow d e^+\nu_e$. a) The Impulse Approximation (IA), where the weak interaction takes place on one proton while the other
proton is just a spectator.  The final state, represented by the grey bubble, is the
resulting neutron and proton being bound as a deuteron. b) A two-nucleon process
where the protons are undergoing a strong interaction at the same time as
the weak interaction (represented by the black circle).}} 
\label{fig1}
\end{figure}

The cross-section for $pp\rightarrow de^+\nu_e$ has been studied extensively
\cite{Sal52,Bli65,Bah69,Gar72,Dau76,Bar79,Gou90,Car91,Kam94,Iva97,Sch98,Par98} and contains two components.  First, an `Impulse Approximation' (IA)
contribution where the weak interaction takes place on a single proton, as shown
in Fig.~\ref{fig1}a).  This comprises more than 95\% of the cross-section at energies
of solar interest and is very well understood because it concerns only the weak
properties of a proton and well-known proton-proton scattering physics.  The second
component, and the remainder of the cross-section, is unconstrained by experiment
and has led theorists to try to improve our understanding of this process.
Conventional potential model calculations will introduce so-called 
meson-exchange currents; essentially two-nucleon effects where the weak 
interaction occurs while
the two nucleons are interacting with each other.   An
example of such a process is shown in Fig.~\ref{fig1}b).  These two-nucleon effects introduce
physics that cannot be constrained directly by $pp$ elastic scattering experiments, and the 
parameters
of this extended physics must be constrained using other methods. Few other methods exist.
%
Neutrino-deuteron scattering could, in principle, constrain the two-nucleon effects.  
Experiments involving neutrino-deuteron scattering are, however, quite difficult. Reactor antineutrino-deuteron experiments can be performed, where the
experimental uncertainties are of order 10-20\%. Nonetheless, they can be used
to infer a constraint on the two-nucleon matrix elements~\cite{bcv}.  
One interesting method is to use tritium $\beta$-decay to constrain the two-nucleon matrix 
elements needed in $pp\rightarrow de^+\nu_e$~\cite{Sch98}. The tritium half-life is well known,
and the same two-nucleon effects occur here as in $pp$ fusion. However, there are 
three nucleons in tritium, hence we cannot be certain that this more complicated structure
does not, in turn, complicate the weak interaction physics further.

If we were to rely purely on theory to predict the two-nucleon component of the $pp$ fusion
cross-section, we still have one more problem.  How do we define the uncertainty in a 
theoretical calculation?  In quantum field theories, such as quantum electrodynamics,
there is a perturbative expansion parameter that dictates the size of contributions at
each order in perturbation theory and thus the size of any neglected remainder.  
The
remainder represents the uncertainty in the theory.  However, conventional nuclear
physics calculations use   Schr\"odinger's equation and nucleon-nucleon potentials to calculate
initial and final-state wavefunctions and, in turn, calculate the matrix elements of
relevant transition operators.  In this approach, no
scheme exists for estimating the size of contributions from unknown or neglected physics.
An alternative to using wavefunctions to study our problem is to use 
an effective field theory (EFT).  An EFT has many benefits, but there
are drawbacks. If the EFT is perturbative (i.e., there is a small expansion
parameter), then one can compute to a specific order in the parameter and then
make direct statements of the scale of theoretical uncertainty of the
problem as determined by the relative size of the next term in the expansion - just as one would
do in quantum electrodynamics.
This is particularly important given that EFTs typically have
an infinite number of terms, and the expansion parameter is required
to ensure that only a finite number of terms are needed for any
given calculation.  In the particular case of nuclear
EFTs, specifically ones where all interactions are of zero-range (i.e., no pions), 
the expansion parameter is denoted by $\epsilon=Q/M_\pi$, where
$Q$ represents a generic momentum in the problem and $M_\pi$ is the pion mass. 
An added benefit of this approach is that
field theories 
pose no problems for conservation laws such as electromagnetic gauge
invariance, which are often difficult to preserve when wavefunctions are used.  
However, EFTs cannot escape the problem 
that all parameters in an EFT must be constrained by experiment, including the various
weak interaction parameters. 

\section{Proton-Proton Fusion}

There have been a number of works dealing with $pp$-fusion in
EFT \cite{KRpp,KRem,BC}, using the EFT without pions developed in 
refs.~\cite{KSW96,K97,vK97,Cohen97,BHvK1,crs}.  As with the traditional 
calculations, the strength of the weak interaction on an interacting 
pair of nucleons is unknown and this ignorance can be specified through one parameter, which
we label $L_{1,A}$~\cite{BC}.  $L_{1,A}$ represents the strength of the interaction shown in
Fig.~\ref{fig1}b). It also appears in the
various $\nu(\overline\nu)-d$ breakup reactions and a precise measurement
of one of those reactions would also constrain $pp$-fusion.  However, 
precise measurements do not exist (but have been proposed~\cite{ORLaND}).  

It is also interesting to consider the inverse of this argument.  A
constraint on $pp$-fusion would lead to a constraint on $\nu(\overline\nu)-d$ 
breakup reactions.  The recent results from the Sudbury Neutrino Observatory (SNO)~\cite{SNO1,SNO2} rely on the reaction $\nu_ed\rightarrow e^-pp$
to detect and extract the solar neutrino flux, hence the theoretical uncertainties in the
cross-section certainly affect their results.
In spite of the importance of SNO, not as much attention has been paid to these breakup reactions,
particularly the pair
\begin{eqnarray}
\nu_ed&&\rightarrow e^-pp\>,\\
\nu_xd&&\rightarrow\nu_xpn\>.
\end{eqnarray}
The first reaction commonly known as the Charged-Current (CC) reaction, is sensitive
to the electron neutrino ($\nu_e$) flux only.  The second Neutral-Current (NC) reaction is equally
sensitive to electron, muon, and tau neutrinos (denoted by $\nu_x$), and provides the key test for neutrino
oscillations.  The CC reaction is very closely related to $pp$-fusion, but the finite
energy/momentum transfers in the CC and NC reactions
make conventional calculations much more difficult than the (relatively) simpler
$pp$-fusion calculation.  However, both
$pp$-fusion and deuteron breakup can be calculated using the same EFT in a 
straightforward manner and the connection between the two is discussed elsewhere~\cite{BC}.
Here, we wish to explore the possibility of using the high-precision 
associated with helioseismology to impose some additional constraints on 
$L_{1,A}$.  In turn, the implications of these constraints can be seen
for $\nu-d$ breakup and for SNO.

Cross-sections of astrophysical interest are written as
\begin{equation}
\sigma(E)={S(E)\over E} \exp(-2\pi\eta)\>,
\end{equation}
separating the nuclear physics ($S(E)$) from the longer range Coulomb
effects (the Gamow factor $\exp(-2\pi\eta)$).  In the particular case
of $pp$-fusion, we are interested in $S_{pp}(0)$.
Adelberger {\it et al.}\/ \cite{Ade98} parameterize $S_{pp}(0)$ as
\begin{equation}
S_{pp}(0)=4.00\times 10^{-25} {\rm MeV\,b} 
\bigg({(ft)_{0^+\rightarrow 0^+}\over 3073\ {\rm s}}\bigg)^{-1}
\bigg({\Lambda^2\over 6.92}\bigg)\bigg({g_A/g_V\over 1.2654}\bigg)^2
\bigg({f_{pp}^R\over 0.144}\bigg)
\bigg({1+\delta\over 1.01}\bigg)^2
\label{adelberger}
\end{equation}
where the parameters are described in ref.~\cite{Ade98}.  Of importance here
are $\Lambda$ and $\delta$.  Their parameterization uses an explicit
separation of IA and two-nucleon effects into $\Lambda$ and $\delta$
respectively.  In this separation the total matrix element for the fusion process
is given by $\Lambda(1+\delta)$.   This separation is artificial and is not appropriate
when a fully self-consistent nuclear physics calculation is performed.
Consequently, we set $\delta=0$ and incorporate all effects directly into $\Lambda$.

In EFT, it was found that \cite{BC}
\begin{equation}
\Lambda=2.58+0.011\left( \frac{L_{1,A}}{\text{1 fm}^{3}}\right)
-0.0003\left( \frac{K_{1,A}}{\text{1 fm}^{5}}\right) \ .  \label{Lambda0N}
\end{equation}
where $L_{1,A}$ parameterizes two-nucleon physics at next-to-leading order (NLO),
and $K_{1,A}$ parameterizes two-nucleon physics at next-to-next-to-next-to-leading
order (NNNLO).  No new parameters appear at the intermediate order (NNLO).
Naively, each order of the expansion will be (naively) 30\%  of the
previous order.  So, a calculation to N$^4$LO should have a precision of 
1\%.  In practice, low-energy calculations of this type are usually more rapidly
convergent.  $L_{1,A}$ and $K_{1,A}$ are both unknown, and need to
be constrained experimentally.
Dimensional analysis as developed in refs. \cite{crs,KSW}
favors
\begin{eqnarray}
\left| L_{1,A}\right| &\approx &\frac{1}{m_{\pi }\left( m_{\pi }-\gamma
\right) ^{2}}\approx 6\text{ fm}^{3}\ ,  \nonumber \\
\left| K_{1,A}\right| &\approx &\frac{1}{m_{\pi }^{2}\left( m_{\pi }-\gamma
\right) ^{3}}\approx 20\text{ fm}^{5}\ .  \label{da}
\end{eqnarray}
It can be shown that $K_{1,A}$ would contribute at an order
much less than 1\%, so we neglect it~\cite{BC}.

\section{Helioseismology and the Standard Solar Model}

The solar models for this investigation were constructed using the Yale
stellar evolution code (YREC) in its non-rotating configuration 
\cite{Gue97,Pin88,Gue92}.
The code implements the most current constitutive physics available.
The equation of state tables are provided by the OPAL group
\cite{Rog86,Rog96}. The opacities in the
interior are interpolated from the OPAL opacity tables \cite{Igl96},
 and the lower temperature opacities, near the surface, are
interpolated from the Alexander and Ferguson \cite{Alx94} tables.  

The nuclear reaction cross-sections have recently been updated (compared to
\cite{Gue97}) to correspond to the latest agreed upon
values set at the 1998 Seattle Workshop on Solar Fusion Reactions.
 The nuclear energy generation routines are nearly identical
to those used by Bahcall {\it et al.}\/ \cite{Bah95} who, themselves, use a variation of YREC for
their solar model calculations. To facilitate quick adjustments of 
S-factors for each reaction, YREC uses Bahcall \cite{Bah89} values
multiplied by a scaling factor $S(0)/S^*(0)$, where $S(0)$ is the revised
S-factor, and $S^*(0)$ is the original Bahcall value \cite{Bah89}. 
For the results presented in this investigation, we
adjust the scaling factor corresponding to the $pp\rightarrow de^+\nu_e$ reaction.  
The models were evolved from the zero age main sequence (ZAMS, the point in the
evolution of the star where the nuclear power first dominates
gravitational power) in 100 equally spaced time steps to 4.55~Gyr. The
meteoritic age of the Sun, that is the age implied from the age of the
oldest meteorites, is 4.52$\pm$0.04 Gyr \cite{Gue89}. Here we use
4.55~Gyr for the
age of the solar model, which lies within the uncertainty range of the meteoritic age, 
to facilitate testing and comparison of our models to solar
models that we and others have calculated in the past. The solar radius, $R_\odot =
6.958\times 10^{10}$~cm, and the solar luminosity, 
$L_\odot = 3.8515\times10^{33}$~erg~s$^{-1}$, are
known to better than $\pm$0.05\%. The mass of the Sun is $1.989\times10^{33}$~g with a relative
uncertainty of $\pm$0.02\% \cite{Coh86}. The mixing length
parameter, a free parameter in the mixing length theory~\footnote[1]{Mixing
length theory describes the efficiency of convection via a single parameter,
the mixing length parameter.  The parameter describes the characteristic 
distance over which a convective element travels before dispersing into its 
surroundings \cite{Gue92}.} used to model convective energy transport, and the helium
abundance, an element whose abundance cannot be  determined
from spectroscopic measurements of the photosphere of the Sun, are
adjusted to produce models that match each other and the above-stated values of
$R_\odot$ and $L_\odot$ to one part 
in $10^6$.

In order for
the surface composition of the (evolved) solar model to match the
observationally determined abundances, we set the initial (ZAMS)
homogeneous composition for hydrogen, helium and metals (A$>$4)
by mass fraction to 0.70633, 0.27367, and 0.02000, respectively. The effects of
gravitational settling of helium and metals, which alters the surface
abundances as the models evolve, are modeled using the diffusion
formulation described by Bahcall et al. \cite{Bah95}. Other modeling details
are as described in \cite{Gue97}. 

To constrain the range for $L_{1,A}$, a relationship between $L_{1,A}$ and
$S_{pp}$ is established.  Neglecting the higher-order effects of $K_{1,A}$
in eq.\ \ref{Lambda0N},
\begin{equation}
\Lambda=2.58+0.011\left( \frac{L_{1,A}}{\text{1 fm}^{3}}\right)
\   \label{Lambda0N2}
\end{equation}
and applying the appropriate values from \cite{Ade98} to eq.\ \ref{adelberger} gives
\begin{equation}
S_{pp} (0)=0.56663\times 10^{-25} (\Lambda^{2}) {\rm MeV\,b} \>. 
\label{spp}
\end{equation}
The combination of eqs.\ \ref{Lambda0N2} and \ref{spp} establishes a relationship between 
$L_{1,A}$ and $S_{pp}$(0), given by
\begin{equation}
S_{pp}(0)=3.7717\times 10^{-25}\bigg(1+8.53\times 10^{-3}\ \bigg({L_{1,A}\over 1\ {\rm fm}^3}\bigg)+1.82\times 10^{-5}\ \bigg({L_{1,A}\over 1\ {\rm fm}^3}\bigg)^2\bigg)\ {\rm MeV\,b}\>.
\end{equation}
Using the above relationship and the current theoretical 
value for $S_{pp}$(0), the best determined value for $L_{1,A}$ may be calculated.
  By varying the $S_{pp}$(0)
value within YREC in accordance with values of $L_{1,A}$, solar models 
are produced corresponding to each variation.  

\begin{figure}[t]
\centerline{{\epsfxsize=5.0in \epsfbox{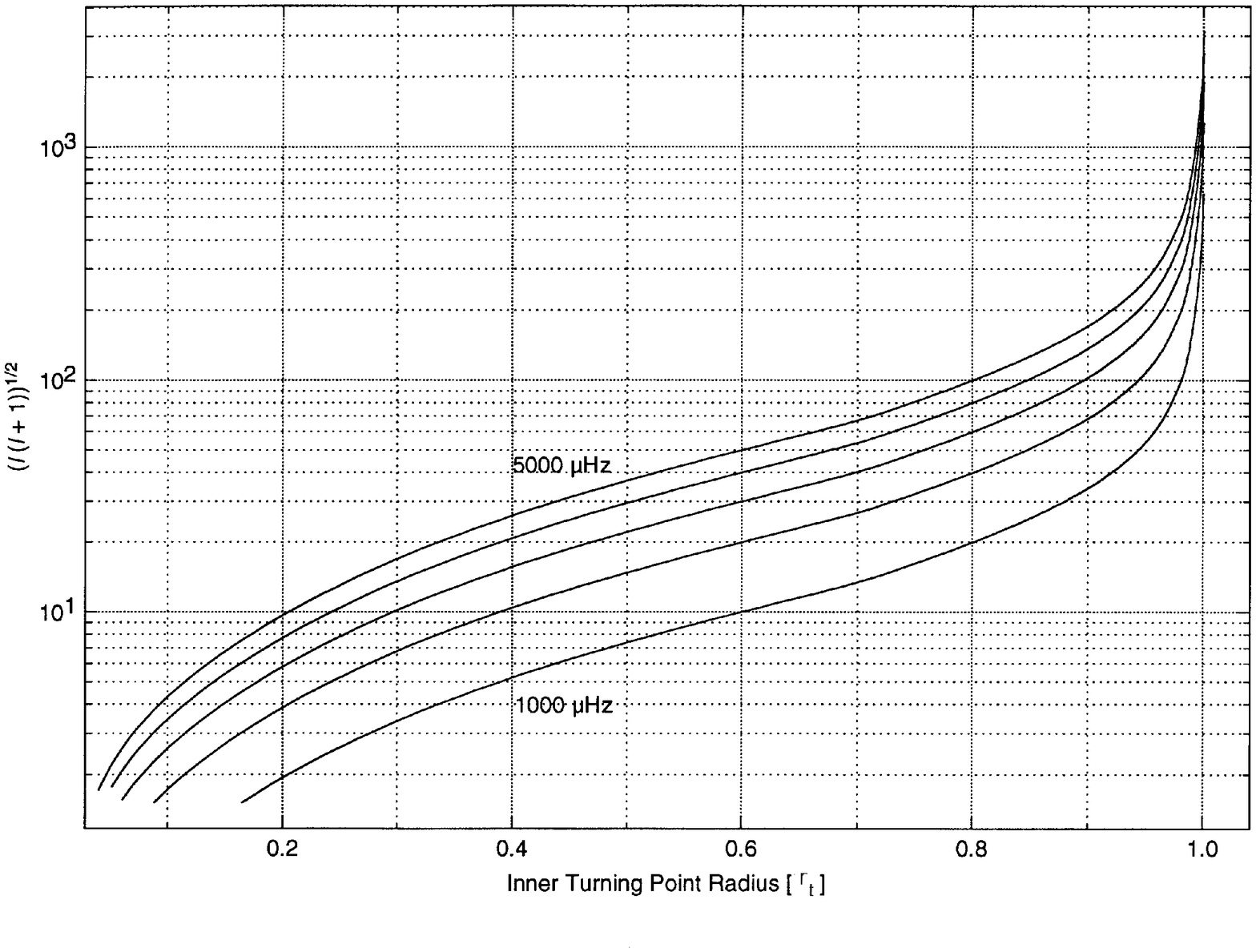}}}
\noindent
\caption{{\it Five curves representing $p$-mode frequencies at 1000 $\mu$Hz,
2000 $\mu$Hz,..., 5000 $\mu$Hz in a plot of the radius fraction of the inner turning point versus azimuthal order.  The inner turning point approaches the core
as $l$ decreases.}} 
\label{unno}
\end{figure}

The resultant models are then read into a nonradial, nonadiabatic
stellar oscillation program \cite{Gue94} where the oscillation
frequencies are calculated. We adopt the usual nomenclature describing
different orders of acoustic oscillations, or $p$-modes, with the radial
order of the mode given by $n$ and the azimuthal order of the mode given
by $l$. Only the lowest $l$-valued $p$-modes (i.e., $l=0,1$) penetrate into the central
nuclear burning regions of the sun. The radius to which a mode penetrates, the 
inner turning point $r_t$, is related to the cyclic frequency $\omega$ and the
sound speed $c(r_t)$ by \cite{unno} 
\begin{equation}
r_t={\sqrt{l(l+1)} c(r_t)\over\omega}\>.
\end{equation}
This behaviour is shown in Fig.~\ref{unno}. 

%
The physics used to model the structure near the surface of the solar model is more uncertain
than the physics of the core.  Indeed, comparisons of $p$-mode frequencies that probe the surface
reveal greater discrepancies between model and observations than in the core.  It is believed
that the mixing length approximation used to model convection is the principal modeling uncertainty.
The frequencies of all observed $p$-modes are affected by this region.
It is possible,
though, to select pairs of $p$-modes that have identical eigenfunction
shape near the surface, and distinct shape in the deep interior, and
use them in combination to cancel out the surface layer effects. This is achieved via the
small frequency spacing, $\delta\nu(n,l)$, which is defined as the frequency difference
between an $\nu(n, l)$ and an $\nu(n-1, l+2)$ $p$-mode. For $l$
= 0 and 1, the $\delta\nu(n,l)$ is well known as one of the most sensitive
helioseismic diagnostics of the central-most regions of the Sun.

\section{Results}

The best value for $L_{1,A}$ is determined by using the current value for
$S_{pp}$(0). From \cite{Ade98}, the current value for $S_{pp}$(0) is
\begin{equation}
S_{pp} (0)=4.00(1\pm0.007^{+0.020}_{-0.011})\times 10^{-25} {\rm MeV\,b} 
\label{spp0}
\end{equation}
Using eqs. \ref{Lambda0N2} and \ref{spp0}, the determined value for the unknown counterterm is found to be, 
$L_{1,A}$ = 7.0 fm$^{3}$.

To determine the range in $L_{1,A}$ using helioseismology, plots of $\delta\nu(n,l)$
versus adiabatic oscillation frequencies are created.  To enhance the analysis, 
$\delta\nu(n,l)$ differences are used instead of $\delta\nu(n,l)$.  These
differences are formed by taking the $\delta\nu(n,l)$ results from the individual 
models and subtracting the observed $\delta\nu(n,l)$ for the Sun; $\delta\nu_{model}(n,l)$
- $\delta\nu_{\odot}(n,l)$, where $\delta\nu_{\odot}(n,l)$ values are calculated from 
observed solar oscillation frequencies.

Low-$l$ small spacings have been published for a number of different high
quality solar $p$-mode observations, including GONG (Global Oscillation
Network Group; \cite{Har96,Chr96})
and BiSON (Birmingham Solar Oscillation Network; \cite{Cha99}).
Here we choose the BiSON data set because the uncertainties are low and
because we have used this data set before and are familiar with it.  

\begin{figure}[t]
\centerline{{\epsfxsize=5.0in \epsfbox{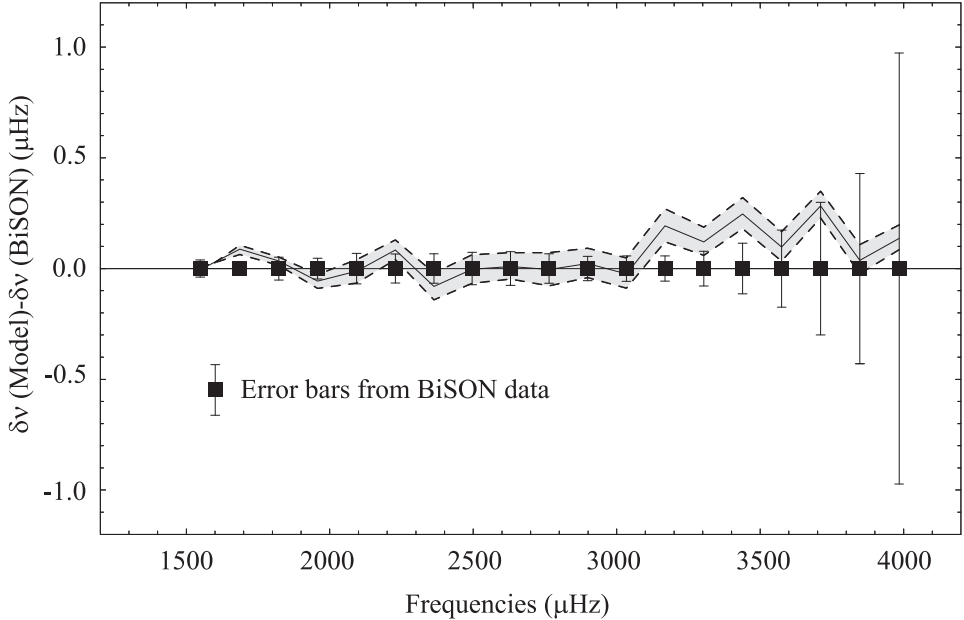}}}
\noindent
\caption{{\it The small spacing frequencies $\delta\nu(n,l)$ for the reference
SSM, shown relative to the observed values from BiSON.  The error bars represent
the uncertainties in the BiSON data.  The discrepancies between the model and
observations are largest at high frequencies, which are most sensitive to surface
effects.  The reference SSM uses a value for $L_{1,A}$ of 7.0~fm$^3$, and
is shown by a solid black line.  The shaded area represents the variation
of $L_{1,A}$ between 1.6 and 12.4~fm$^3$.}}
\label{SSM}
\end{figure}

A plot of $\delta\nu(n,l)$ differences versus adiabatic frequencies for the 
reference SSM is shown in Fig.~\ref{SSM}.  In Fig.~\ref{SSM}, the thick solid line represents
the reference SSM model result.  The  error bars correspond to the uncertainties stated for
the observed BiSON results for the Sun \cite{Cha99}. 
%

The low-$l$ small spacing frequencies of our reference SSM lie almost
entirely within the uncertainty range of the observationally derived
small spacings. This suggests that the interior structure of our model
is a very close fit to the actual sun. Which, in turn suggests that the
physics specific to this region are accurate, including the age
estimate, the chemical abundance and diffusion processes, the
opacities, the equation of state, and the nuclear reaction rates. That
is, no modifications to the current SSM are required to make it agree
with the most accurate $p$-mode data we currently have available to us.
It is comforting that the simplest, non-ad-hoc, physics has produced an
accurate match to the observed $p$-mode data.

The analysis described above was repeated for each, nonstandard, model corresponding to a
variation of $L_{1,A}$ with respect to the reference SSM.  
A quantitative estimate of the range of $L_{1,A}$ is established by comparing
the distribution of the frequency results for various $L_{1,A}$ with respect to the observed
BiSON error bars.  To extract this quantitative result, the frequency range of $\sim$2500 to 
$\sim$3000 $\mu$Hz was studied since this frequency range for the SSM showed an excellent fit to
the observed BiSON data.

For each model results, average $\delta\nu(n,l)$ difference values were calculated within the
frequency range of $\sim$2500 to $\sim$3000 $\mu$Hz.  The average observed error within this 
frequency range was calculated to be $\pm$0.066 $\mu$Hz.  Average $\delta\nu(n,l)$
differences are plotted versus the corresponding $L_{1,A}$ value for both the upper and lower
range.  These plots show a linear relation between the average $\delta\nu(n,l)$ differences
and $L_{1,A}$.  From this relation and the average observed error values, the values of 
$L_{1,A}$ corresponding to the upper and lower limits of the error bars are calculated.
The upper limit is determined to be 12.4 fm$^3$.  The lower limit is determined to be 
1.6 fm$^3$. The range of allowed values is shown as the shaded area in Fig.~\ref{SSM}.
Therefore the value and effective range for $L_{1,A}$ is determined to be
\begin{equation}
L_{1,A} = 7.0\pm5.4\ {\rm fm}^3
\end{equation}

A further result from the determined SSM is the calculation of the theoretical total
$^8B$ neutrino flux of $4.93\times10^6$ cm$^{-2}$s$^{-1}$.  This result is consistent with the total
observed $^8B$ neutrino flux from SNO of $(5.09 ^{+0.44}_{-0.43}{\rm (stat)} ^{+0.46}_{-0.43}{\rm (syst)})\times10^6$ cm$^{-2}$s$^{-1}$ \cite{SNO2}.

We have tacitly assumed that the uncertainty in the model-computed $p$-mode frequencies is
comparable to, or less than, the error bars associated with the observed frequencies. This is not
too unreasonable since the model-computed frequencies do lie within the uncertainty error
bars of the observations.  In order to more formally justify this assumption, we must carry out an
extensive uncertainty analysis of the solar model physics and their impact on the computed $p$-mode
frequencies.  This important task, upon which all similar helioseismic tests depend, is in progress and
will be reported at a later time.

\section{Conclusions}

Using the relationship between effective field theory, nuclear cross-sections and the accuracy of helioseismology,
the unknown counterterm, $L_{1,A}$, is determined to have a value of 7.0 fm$^3$ with a range of 
1.6 to 12.4 fm$^3$.  This result for $L_{1,A}$ is consistent with the theoretical value 
determined by Butler $\&$ Chen \cite{BCK} using dimensional analysis of $L_{1,A}$ $\approx$ 6 fm$^3$.
The result and range is also consistent with other evaluations of $L_{1,A}$ which include
5.6$\pm$2.0 fm$^3$ \cite{NSGK} and 6.5$\pm$2.4 fm$^3$ \cite{Sch98}.

The standard solar model determined for this work shows remarkably excellent agreement with the latest BiSON oscillation
frequency observations.  Theoretical total fluxes for $^8B$ neutrinos from the standard solar model result also show 
excellent agreement with total observed $^8B$ fluxes reported from the Sudbury Neutrino Observatory.

\section{Acknowledgments}
The authors would like to thank the Natural Sciences and Engineering Research Council (NSERC) of Canada
for its support of this research.

\end{document}